\documentclass[nofootinbib,twocolumn]{revtex4}
\usepackage{graphicx}
\usepackage{amssymb}
\usepackage{amsmath}
\usepackage{bm}
\usepackage{hyperref}
\begin{document}

\title{Newtonian View of General Relativistic Stars}

\author{A. M. Oliveira}\email{adriano.oliveira@ifes.edu.br}
\affiliation{Grupo de Ci\^encias Ambientais e Recursos Naturais, Instituto Federal do Esp\'irito Santo (IFES), Guarapari, Brazil}
\author{H. E. S. Velten}\email{velten@pq.cnpq.br}
\author{J. C. Fabris}\email{fabris@pq.cnpq.br}
\affiliation{Departamento de F\'isica, Universidade Federal do Esp\'{\i}rito Santo (UFES), Vit\'oria, Brazil}
\author{I. G. Salako}\email{ines.salako@imsp-uac.org}
\affiliation{Institut de Math\'ematiques et de Sciences Physiques (IMSP), Porto-Novo, B\'enin}

\begin{abstract} 
Although general relativistic cosmological solutions, even in the presence of pressure, can be mimicked by using neo-Newtonian hydrodynamics, it is not clear whether there exists the same Newtonian correspondence for spherical static configurations. General relativity solutions for stars are known as the Tolman-Oppenheimer-Volkoff (TOV) equations. On the other hand, the Newtonian description does not take into account the total pressure effects and therefore can not be used in strong field regimes. We discuss how to incorporate pressure in the stellar equilibrium equations within the neo-Newtonian framework. We compare the Newtonian, neo-Newtonian and the full relativistic theory by solving the equilibrium equations for both three approaches and calculating the mass-radius diagrams for some simple neutron stars equation of state.\\


\end{abstract}

\maketitle

\section{Introduction}

General relativity is the usual theory for dealing with gravitation phenomena \cite{weinberg}. Its building blocks, like the equivalence principle and predictions for the trajectories of planets and light in the solar system, have passed for the most different tests. So far, we still do not have any clear evidence against GR though there are, of course, many alternative theories like Brans-Dicke \cite{BD}, $f(R)$ \cite{fR} and others \cite{MG}. 

The application of GR to expanding backgrounds gives rise to cosmological solutions. For expanding, homogeneous and isotropic configurations we obtain from GR the well know Friedmann equations for the evolution of the cosmological scalar factor $a(t)$. These cosmological solutions were obtained in 1922, a few years after Einstein presented the GR field equations. 

A quite interesting fact is that in the 1930s Milne and McCrea started to develop a Newtonian description for cosmology \cite{Milne}. From their work, it has appeared the so called Newtonian Cosmology from which one obtains the same relativistic solutions for matter dominated universes, i.e., Newtonian cosmology matches GR for presureless fluids.

With the works of Milne in the 1950s and Harrisson in 1965, pressure effects were correctly incorporated into Newtonian cosmological solutions. This new approach became famous as the neo-Newtonian cosmology and all the GR Friedmann solutions could now be obtained \cite{McCrea1951, Harrison1965}. Therefore, via the neo-Newtonian formulation, cosmology can be studied with a much simpler formalism than the full GR theory.

Cosmology represents only one type of solutions which can be obtained from gravitational relativistic theories. Black holes configurations and gravitational waves are also examples of solutions. For static configurations GR is able to describe the structure of objects like stars. The GR solutions for isotropic stars are known as the Tolmann-Oppeinheimer-Volkoff (TOV) equations from which we can solve the equilibrium configuration of the stellar interior \cite{OV, Tolman}.

In general, nuclear reactions within the stellar interior induce energy flow via radiative convection. However, since for compact stars the nuclear timescale is much larger than the thermal and dynamical timescales one can assume the hydrostatic equilibrium for most of the star lifetime. In a first approximation, hydrostatic equilibrium in stars can be studied with Newtonian mechanics. From this approach one obtains the Lane-Emden equation \cite{Chandrasekhar} which is basically the Newtonian limit of the TOV system when pressure does not source gravitational effects in the stellar interior. Of course, since the pressure effects, mainly in extreme relativistic stars, are very important for the stellar properties, many systems in nature can not be described via the Newtonian equations. One important example is a neutron star. Hence, one may wonder: is it possible to reproduce the full TOV solutions using neo-Newtonian hydrodynamics? In other words, is there the same equivalence between GR and the neo-Newtonian framework for a static configuration? This is the main goal of this paper. See also Ref. \cite{pedraza} for interesting related investigations.

In the next section we briefly review the neo-Newtonian cosmology. This review will be useful to present the neo-Newtonian hydrodynamics which will be important for the rest of the paper. A crucial aspect of this discussion concerns the equivalence principle which
 will also be addressed here through the different possible definitions of mass. In section III we briefly review the Newtonian hydrostatic equilibrium equation and the Tolman-Oppenheimer-Volkoff (full relativistic) equation for star. Then we propose the implementation of the neo-Newtonian hydrodynamics for stellar configurations. We show in the section IV numerical results for the mass-radius diagram for neutron stars adopting well know equations of state (EoS) given in the literature. We conclude in the final section.

\section{Newtonian Hydrodynamics in an Expanding Background: Cosmology}

The purpose of this section is to present the neo-Newtonian hydrodynamics applied to cosmology. However, let us start with the standard case of Newtonian equations.
The basic equations of Newtonian hydrodynamics for an inviscid perfect fluid are the following:
\begin{eqnarray} 
\dot{\rho} + \nabla\cdot(\rho\vec{v})  &=& 0, \label{eq100'}\\
\rho \frac{d \vec{v}}{dt} \equiv  \rho [\dot{\vec{v}}+
(\vec{v}\cdot\nabla)\vec{v} ] &=& -\nabla p, \label{eq200'}
\end{eqnarray}
where $\rho$ is the fluid density, $p$ its pressure and $\vec v$ its the velocity field. The dot means derivative with respect to the cosmic time $t$, i.e., $\dot{a}=\partial_t a$.

The above system of equations becomes suitable to study cosmology adopting the velocity field $\vec{v}=H (t) \vec{r}$ (Hubble's law) where $H(t)=\frac{\dot{a}(t)}{a(t)}$, being $a(t)$ the scale factor. It is worth noting the trivial solution for the continuity Eq. (\ref{eq100'}) $\rho(a)=\rho_0/a^3$, where the today's scale factor $a_0=1$ gives the today's density of the fluid $\rho_0$. 

Gravitational interaction is coupled into Euler's equation (\ref{eq200'}) as
\begin{equation}\label{euler1} \dot{\vec{v}}+
(\vec{v}\cdot\nabla)\vec{v}  = - \frac{ \nabla p}{\rho} - \nabla \Psi ,
\end{equation}
where the gravitational potential $\Psi$ obeys the Poisson equation
\begin{equation}
\label{poisson1}\nabla^{2} \Psi = 4 \pi G \rho.
\end{equation}

Eqs (\ref{eq100'}) and (\ref{euler1}) provide a fluid picture of the cosmic medium which is gravitationally self-interacting via the Poisson Eq. (\ref{poisson1}). In the Newtonian cosmology the Friedmann equations read

\begin{equation}
\frac{\dot{a}^2}{a^2}+\frac{\left(-2E\right)}{a^2}=\frac{8\pi G}{3} \rho \hspace{0.5cm}{\rm and}\hspace{0.5cm}\dot{H}+H^2=-\frac{4\pi G}{3}\rho,
\end{equation}
where $E$ is a constant of integration associated to the energy of system. The pressure is not dynamically relevant for the homogeneous and isotropic background. With Newtonian cosmology it is not possible to model a radiation dominated phase or even to
study a late time dark energy dominated epoch. This approach is restricted to a description of the Einstein-de Sitter universe.

\subsection{Including Pressure: The neo-Newtonian Cosmology}

A simple way to include the pressure effects and, at the same time, keep the simplicity of the Newtonian physics is the use of the neo-Newtonian equations developed during the 1950s by McCrea \cite{McCrea1951} and by Harrison in the 1960s \cite{Harrison1965}. Later, during the 1990s, an important analysis concerning the perturbative behavior of the neo-Newtonian equations has helped to set the final form for the fluid equations in this approach \cite{Lima1997} (see also \cite{RRRR2003}). This set of equations reads,
\begin{equation}\label{continuity2}
\dot{ \rho}+ \nabla \cdot \left( \rho \vec{v} \right)+p\nabla \cdot \vec{v}=0\,,
\end{equation}
\begin{equation}\label{euler2}
\dot{ \vec{v}} + (\vec{v} \cdot  \nabla )\vec{v} = - \nabla \Psi   - \frac{ \nabla  p}{\rho + p}\,,
\end{equation}
\begin{equation}
\label{poisson2}\nabla^{2} \Psi = 4 \pi G \left(\rho + 3 p\right).
\end{equation}
Combining Eqs.~(\ref{continuity2}), (\ref{euler2}) and (\ref{poisson2}) one obtains equations for the expansion of the homogeneous and isotropic background that are exactly the same as the relativistic Friedmann equations,
\begin{eqnarray}
\frac{\dot{a}^2}{a^2}+\frac{\left(-2E\right)}{a^2}=\frac{8\pi G}{3} \rho ,\\
\dot{H}+H^2=-\frac{4\pi G}{3}(\rho+3p).
\end{eqnarray}
The main idea behind the neo-Newtonian formalism relies on the following substitutions:
Firstly, it is necessary to redefine the concept of inertial and passive gravitational mass density. With the redefinition
\begin{equation}\label{rhoi}
\rho_i \rightarrow \rho +p,
\end{equation}
we rewrite the continuity and the Euler equation.

The second step is the interpretation of the active gravitational mass density i.e., the density that source the gravitational field. Hence the following redefinition 
\begin{equation}\label{rhog}
\rho_g \rightarrow \rho + 3p,
\end{equation}
which is related to the trace of the energy-momentum tensor, will become the source of the Poisson equation.

\section{Newtonian, relativistic and neo-Newtonian formulations for static configurations: stars}

Although very simple, spherically symmetric and static geometries are perfect for studying astrophysical objects like stars. In this section we start reviewing the Newtonian and general relativistic formulations. They will be useful for a comparison with the neo-Newtonian approach to stellar hydrodynamical equilibrium we present at the end of the section. 

The Newtonian hydrostatic equilibirum equations are widely derived in the literature \cite{weinberg}. They read

\begin{equation}\label{Nequi}
\frac{dp}{dr}=-\frac{G \rho M(r)}{r^2},
\end{equation}
and
\begin{equation}\label{Ndmdr}
\frac{dM(r)}{dr}=4\pi r^2 \rho.
\end{equation}

In principle, the system (\ref{Nequi}-\ref{Ndmdr}) can be solved numerically since an appropriate EoS $p \equiv p(\rho)$ is provided.

On the other hand, there is a similar set of equations when General Relativity is adopted. They are known as the Tolman-Oppeinheimer-Volkoff (TOV) equation,
\begin{equation}\label{TOV}
\frac{dp}{dr}=-\frac{G\rho M(r)}{r^2}\frac{\left(1+\frac{p}{\rho}\right)\left(1+\frac{4\pi r^3 p}{M(r)}\right)}{\left(1-\frac{2G M(r)}{r}\right)}.
\end{equation}
Only with $p\ll\rho$, Eq. (\ref{Nequi}) can not be obtained because of the Schwarzschild-like correction in the denominator (\ref{TOV}). After restoring unities the correct Newtonian limit occurs for $c \rightarrow \infty$.

The TOV equation is coupled to the mass definition
\begin{equation}\label{Rdmdr}
\frac{dM(r)}{dr}=4 \pi r^2 \rho.
\end{equation}
which has the same form as its Newtonian counterpart (\ref{Ndmdr}).

Therefore, the full equilibrium configuration for stellar systems is obtained by solving the system (\ref{TOV}) sourced by the differential relation (\ref{Rdmdr}).

Now we derive the neo-Newtonian correspondence for the hydrostatic equilibrium condition.  In addition to the modified Poisson Eq. (\ref{poisson2}) one finds
\begin{equation}\label{nNequi}
\frac{1}{\rho_i}\frac{dp}{dr}=-\frac{G}{r^2}\int^{r}_{0} 4\pi r^{\prime 2} (\rho+3p) dr^{\prime}.
\end{equation}

Equation (\ref{nNequi}) with the proper neo-Newtonian identifications for inertial/passive-gravitational mass densities results in
\begin{equation}\label{nNTOV}
\frac{dp}{dr}=-\frac{G(\rho+p)}{r^2}\tilde{M}(r),
\end{equation}
where one also uses the relation
\begin{equation} \label{nNdmdr}
\frac{d\,\tilde{M} (r)}{dr}= 4 \pi r^2 (\rho +3 p).
\end{equation}

Note that the above relation differs from the standard definition for the mass of an object
\begin{equation}\label{MneoNew}
M=\int^R_0 4 \pi r^2 \rho dr.
\end{equation}
 
Hence, Eqs (\ref{nNTOV}) and (\ref{nNdmdr}) represent the neo-Newtonian version of the TOV system (\ref{Rdmdr}) and (\ref{TOV}).

\section{Numerical Results}

We will solve now the differential equations for the internal structure of stars. Our goal is to compare the solution for the Newtonian formulation (\ref{Nequi}) and (\ref{Ndmdr}); the full relativistic TOV equations which is (\ref{TOV}) sourced by the differential relation (\ref{Rdmdr}) and the neo-Newtonian Eqs. (\ref{nNTOV}) and (\ref{nNdmdr}). 

White dwarfs are objects in which the simple Newtonian formalism works quite well. Therefore, the comparison we propose here seems to be meanless if using such Newtonian stars. On the other hand, neutron stars are perfect laboratories for testing the relativistic corrections incorporated by the TOV equations. 

After specifying the EoS $p(\rho)$ for the stellar interior, the solution of the equilibrium equations will also depends on specifying the central stellar pressure $p(r=0)=p_0$. Given the $p_0$ value one obtains the total mass and radius of the star. Unfortunately the inner composition of neutron stars are not well understood and the correct neutron star EoS is still unknown. The typical densities found in such objects are of order or greater than the nuclear saturation density where the common way to obtain the EoS of nuclear matter is via microscopic many-body calculations based on phenomenological relativistic mean-field theories and nucleon-nucleon potentials. Usually, either variational or Monte-Carlo techniques are used (see \cite{page, Lattimer} for a review).

One possibility is that precise observations of spin rate, mass and radius of many different stars can lead to the reconstruction of the neutron star properties, i.e., one could indirectly obtain $p(\rho)$. Remarkably, recent observations of very massive neutron star, with mass as high as $1.97 \pm 0.04\, M_{\odot}$ \cite{Mass197}, may place upper limits to thermodynamics quantities like energy density, pressure, baryon number density and chemical potential \cite{NeutronStarObsEoS}. 

There are in the literature many proposals for the nuclear EoS of neutron stars \cite{MIT, pal6, Sly, AP14, FPS, BBB2}. Also, some fits provide a unified EoS for the neutron star interior \cite{analytical}. Here, since our focus relies on the gravity approach we will adopt simple configurations usually adopted as a first contact with neutron stars physics and which can be found in standard textbooks. 

Let us firstly adopt a pure neutron fluid using the Fermi gas model as the Oppenheimer-Volkoff model \cite{OV}. This configuration is, strictly speaking, unrealistic since actual neutron stars contain small fractions of protons and electrons which avoid the neutron decay via weak interactions. Also, a pure neutron configuration fails in counting for the nucleon-nucleon interactions which are important to the energy density. However, we use it as a first analysis of neutron stars. A simple introduction to such EoS can be found in Refs. \cite{weinberg, Silbar}. It corresponds to a polytrope $p\sim \rho^{\gamma}$ with $\gamma=5/3$. We use in this work
\begin{equation}\label{EOS1}
\frac{\bar{\rho}(\bar{p})}{c^2}=\bar{K}^{-1}\,\bar{p}^{3/5},
\end{equation}
where the coefficient of (\ref{EOS1}) is fitted in Ref. \cite{Silbar} as $\bar{K}=1.914$. The bar over the density and pressure makes them dimensionless quantities via the transformations $p=\epsilon_0 \bar{p}$ and $\rho=\epsilon_0 \bar{\rho}$, where $\epsilon_0=1.603$~x~$10^{38}ergs/cm^3$.

For this EoS we plot the mass-radius diagram in Fig.~1. It shows the stable configurations for different values of the central pressure $p(r=0)=p_0$. Low-mass/large-radius solutions correspond to small starting values of $p_0$. We plot with the solid line the TOV solution. The neo-Newtonian (solid-red) and the Newtonian (short-dashed) solutions are also shown. Concerning the TOV solution and the Newtonian one, there is a reliable qualitative agreement of these results. As expected, the relativistic solution presents a maximum (here $M_{max}\sim0.95M\odot$ at $R\sim 8 km$) while for the Newtonian case the larger the mass, the smaller the radius. 

The quantitative equivalence between the TOV and the neo-Newtonian models does not happen. More massive configurations are allowed in the neo-Newtonian context. However, a remarkable result is that the neo-Newtonian solution also presents a maximum mass which is a typical relativistic aspect. This shows that it can be used as a first approximation to the problem.

Performing a quantitative comparison between the Newtonian and the neo-Newtonian cases, although the existence of the maximum, there is indeed, the indication that by adding the pressure to the Newtonian formalism one obtains more compact stars corroborating the relativistic intuition that pressure ``amplify'' gravity effects. 

\begin{figure}\label{fig1}
\begin{center}
\includegraphics[width=0.38\textwidth]{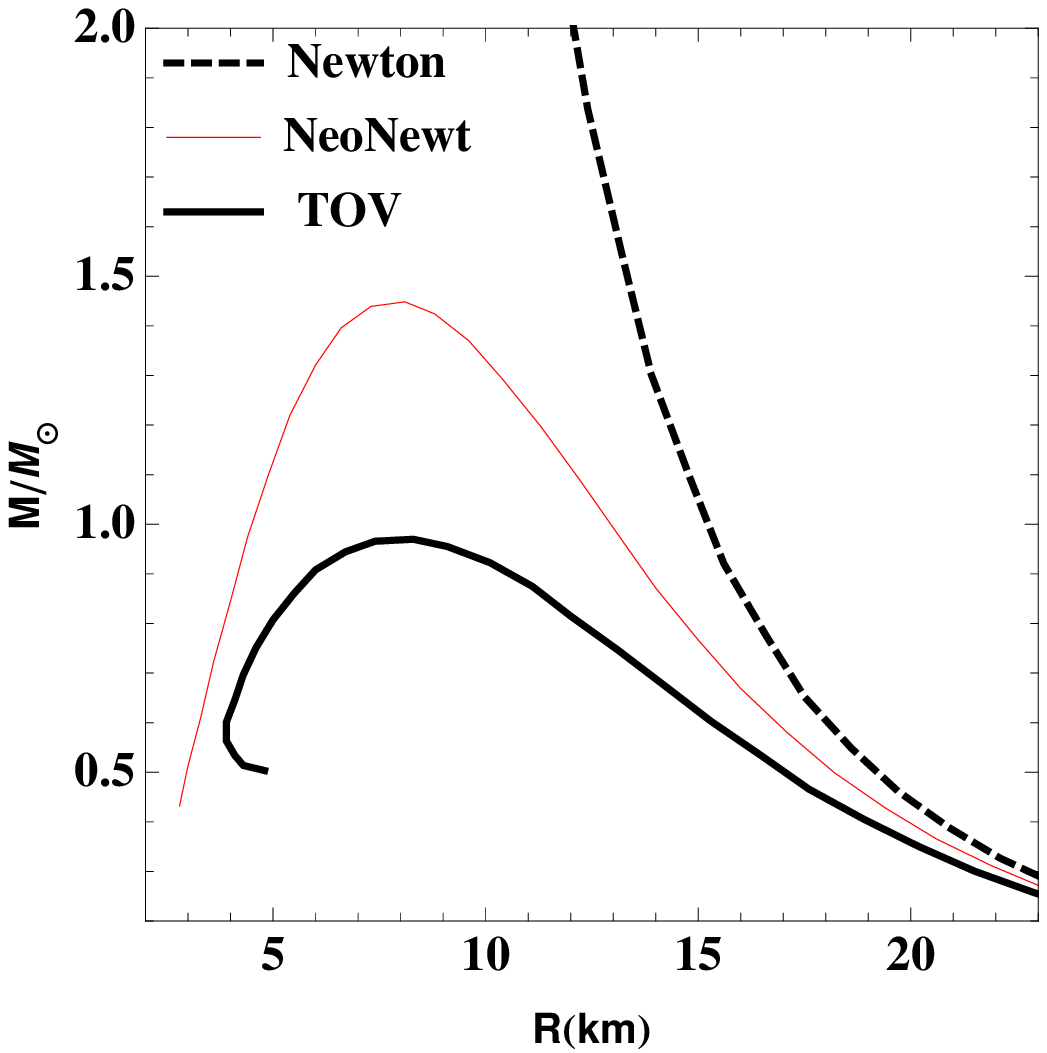}
\caption{Mass-radius contours for a pure neutron star (without interactions) using a Fermi gas equation of state.}
\end{center}
\end{figure}

In Fig.~1 we have calculated the mass of the neo-Newtonian case using the definition (\ref{MneoNew}) which is the same as the other cases and therefore allows us a proper comparison of the results. However, within the neo-Newtonian formulation there is some freedom in defining the meaning of mass. For example, the quantity $\tilde{M}$ (\ref{nNdmdr}) could be used instead. Also, as discussed previously, one has to take care with the definitions of inertial, passive gravitational and active gravitational masses. Hence, being the actual definition of mass in the neo-Newtonian formalism a tricky issue, we plot in Fig.~2 the mass diagrams with a comparison of the possible interpretations of mass within the neo-Newtonian framework. The curve $m_0$ is the same as shown previously in Fig.~1. The quantity $\tilde{M}$ is represented by the long dashed line in these plots. We call it $m_3$, i.e., $m_3\sim \int (\rho+3p) r^2 dr$. Also, we display the case $m_1 \sim \int (\rho+p) r^2 dr$. 

\begin{figure}\label{fig2}
\begin{center}
\includegraphics[width=0.38\textwidth]{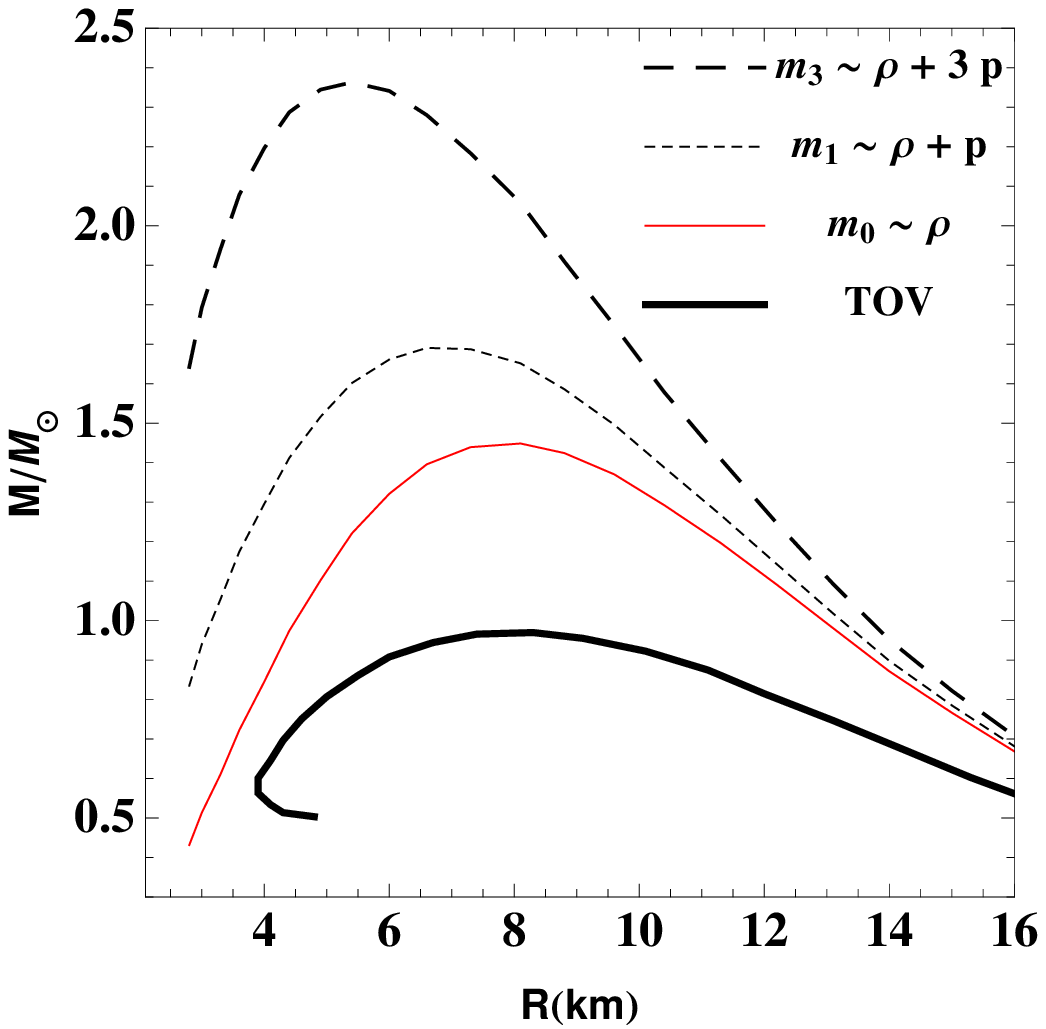}
\caption{Mass-radius contours for a pure neutron star (without interactions) using a Fermi gas equation of state.}
\end{center}
\end{figure}

Now we repeat the above comparison between the three approaches but using a more realistic EoS. We assume a pure neutron star but with nucleon-nucleon interactions. We follow the Prakash method for building such interacting Fermi gas model EoS \cite{prakash}. This equation assumes the polytropic form 
\begin{equation}
\frac{\rho(\bar{p})}{c^2}=(\kappa_0\epsilon_0)^{-1/2}\bar{p}^{1/2},
\end{equation}
where now $\epsilon_0=m^4_n c^5 / 3 \pi^3 \hbar^3$ and the nuclear (in)compressibility is $\kappa_0=363~ MeV$.

Solving again the equilibrium equations, we show in Fig.~3 the mass-radius contours. Here, the Newtonian equations do not apply. Together with the TOV (black solid) and the standard  neo-Newtonian (red solid line) result, we add here the other options for calculating the mass in the neo-Newtonian case (dashed lines). Here again, independently on how we compute the neo-Newtonian mass, it does not agree with the full relativistic solution.

The effect of the nucleons interactions is quite remarkable since the allowed maximum masses are now much larger than the situation in Fig.~1. The TOV maximum mass now $M\sim 2.3~M_{\odot}$ (at $R=13.5~km$) is a reflection of the large value of the nuclear (in)compressibility $\kappa_0=363~ MeV$ used in this analysis. 

In any of the cases studied here, the neo-Newtonian model was not able to correctly, i.e., quantitatively, describe the general relativistic predictions. Therefore, although this formalism represents a reliable tool for studying cosmology, the same does not seem to happen for stellar configurations.

\begin{figure}\label{fig3}
\begin{center}
\includegraphics[width=0.38\textwidth]{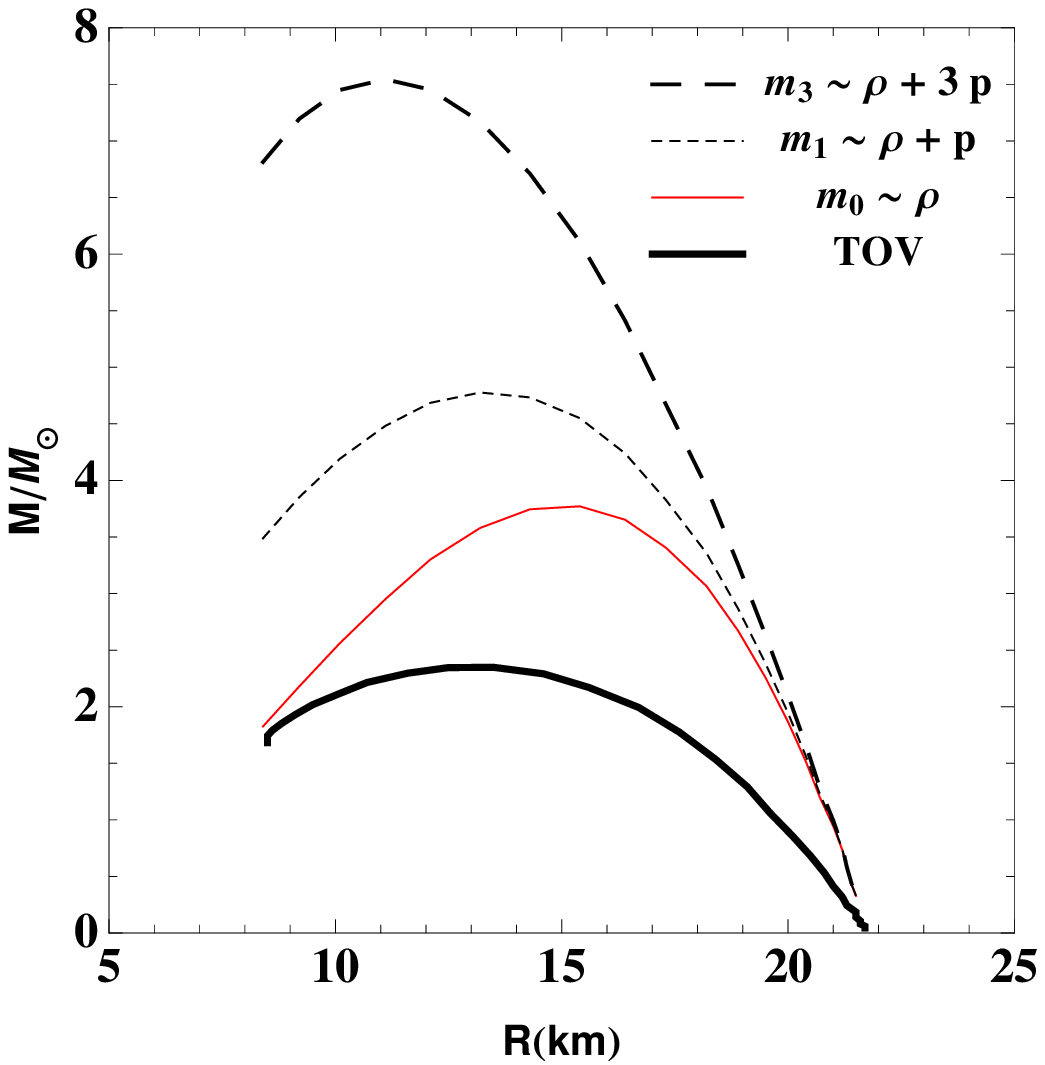}
\caption{Mass-radius contours for a pure neutron star (with nucleon-nucleon interactions) using a Fermi gas equation of state.}
\end{center}
\end{figure}

\section{Conclusions}

It is a remarkable fact that the general relativity predictions for expanding backgrounds can be mimicked by simple Newtonian models. The dynamics of an expanding Newtonian sphere is the same as the relativistic Friedmann-Lemaitre-Robertson-Walker universe dominated by matter, i.e., the Einstein-de Sitter universe. Although the Newtonian cosmology fails in describing epochs of the universe where the pressure is relevant, the neo-Newtonian cosmology was developed to fill this gap.

The question we addressed in this work was to understand whether such exact correspondence between the neo-Newtonian model and the full relativistic theory remains the same for static configurations. Hence, neutron stars seem to be the perfect laboratory for testing our proposal. 

After reviewing the standard Newtonian and the TOV equations, we developed the neo-Newtonian version for the hydrostatic equilibrium in static spherical configurations. The neo-Newtoniam formalism for stars has a structure which looks like the TOV one but without the Schwarzshild-like correction in the denominator of Eq.~(\ref{TOV}). 

We probed both the Newtonian, neo-Newtonian and the relativistic approaches by calculating their equilibrium configurations for some specific neutron stars equation of states. Firstly we used the Fermi gas approximation for a pure neutron star (the classical Oppenheimer-Volkoff) model. The results were shown in Figs.~1 and 2. In general, a clear difference between the Newtonian and the relativistic theory is the existence of a maximum mass in the latter. As one can see in Fig.~1, although the neo-Newtonian result is not the same as the relativistic one, it presents a maximum mass. This result is a remarkable aspect of the neo-Newtonian formalism. However, the neo-Newtonian value for the maximum mass $M_{nN}\sim 1.45 M_{\odot}$ (see Fig.~1) is much larger than we could expect for this EoS from simple analytical arguments \cite{nauenberg}. Therefore, the neo-Newtonian method seems to somehow overestimate the stellar maximum mass.

We also used a more realistic EoS for neutron stars where the interaction nucleon-nucleon is allowed. Concerning the comparison between the fluid models used here, the same qualitative results obtained in the first analysis are kept.

It is therefore shown that the neo-Newtonian hydrodynamics formulated to match GR predictions for expanding background reproduces qualitatively the relativistic effects when applied to spherically symmetric configurations but does not fit quantitatively the GR results in this situation.

\textbf{Acknowledgement}: We are thankful to J. A. de Freitas Pacheco for a critical reading and suggestions to the manuscript. We thank CNPq (Brazil) and FAPES (Brazil) for partial financial support. I.G. Salako also thanks ICTP/IMSP (Benin) for partial support.

\appendix
\section{About the Stellar Virial Theorem}

Here, we provide a complementary discussion about stellar configurations by studying the proper modifications to the virial theorem. We keep here standard assumptions like hydrostatic equilibrium and that stars are made up of ideal gases.

The total internal energy of a star can be written as
\begin{equation}\label{Ei}
E_i = \int_0^{M_{T}} u dM = \int_0^{M_{T}} \frac{3}{2}\frac{KT}{\mu} dM = \int_0^{M_{T}} \frac{3}{2}\frac{p}{\rho} dM.
\end{equation}
where we have used the ideal gas EoS, $P=\rho N_A KT / \mu$. The above result forms one of contributions to the virial theorem. The other one comes from the hydrostatic equilibrium. For example, if we multiple (\ref{Nequi}) by $4\pi r^3 dr$ and integrating over the entire star we find
\begin{equation}\label{Eg}
\int^R_0 12 \pi r^2 p dr = - E_g,
\end{equation}
where, as usual, $E_g = - \int_0^R G M(r)\rho 4 \pi r dr $. Since $dM=4\pi r^2 \rho dr$ and having in mind (\ref{Ei}), it results in the standard result \cite{Chandrasekhar}
\begin{equation}\label{VirialNew}
2 E_i + E_g=0.
\end{equation}

Now, let us follow these same steps but introducing the neo-Newtonian concepts. First of all, note that since the computation of the internal energy $E_i$ involves microscopic/thermodynamic physics this quantity will not be changed if we work with either modified kinematical or gravitational concepts. Hence, we will assume that the internal energy (\ref{Ei}) remains the same. 

The neo-Newtonian formalism involves a different definition for the mass. It is worth noting that the definition of $d\tilde{M}$ (\ref{nNdmdr}) can be integrated and written as
\begin{equation}
\tilde{M}-M=\int^R_0 12 \pi r^2 p dr=2 \frac{E_i}{c^2},
\end{equation}
where $M$ obeys to (\ref{Ndmdr}). Note we have recovered the proper units by including the speed of light $c$. Taking the limit $c \rightarrow \infty$ we see that $\tilde{M}=M$ and therefore the Newtonian result is recovered.

We take now the neo-Newtonian hydrostatic equilibrium (\ref{nNTOV}) and proceed as above multiplying it by $4 \pi r^3 dr$ and integrating over the entire star. We find out a very similar virial configuration
\begin{equation}
2 E_i + \tilde{E}_g=0,
\end{equation}
but the modified gravitational energy $\tilde{E}_g$ here reads
\begin{equation} \label{Etilg}
\tilde{E}_g = E_g^{\rho \rho} + 4 E_g^{\rho p} + 3E_g^{pp}.
\end{equation}
The quantities $E^{\rho p}$ and $E^{pp}$ are the new contributions to the virial equilibrium due to the pressure effects. The contributions to (\ref{Etilg}) are defined here using a simplified notation
\begin{equation}
E_g^{XY}= - \int^R_0\frac{G}{r} \left(\frac{4 \pi r^3}{3}\right) X Y 4 \pi r^2 dr,
\end{equation}
where $X$ and $Y$ can be either the density $\rho$ or the pressure $p$.

The combination $X=Y=\rho$ results in the standard Newtonian gravitational energy as used in (\ref{Eg}). The quantity $E_g^{\rho p}$ results from adopting $X=\rho$ and $Y=p$. There is also a pure pressure square term contribution $E^{pp}$ which is calculated adopting $X=Y=p$.

\end{document}